\documentclass[12pt]{article}
\usepackage{graphicx}
\usepackage{amsmath}

\begin{document}

\title{{\Large Green functions of the Dirac equation with magnetic-solenoid field}}
\author{S.P. Gavrilov\thanks{%
Dept. Fisica e Quimica, UNESP, Campus de Guaratingueta, Brazil; On leave
from Tomsk State Pedagogical University, 634041 Russia; e-mail:
gavrilovsp@hotmail.com}, D.M. Gitman\thanks{%
e-mail: gitman@dfn.if.usp.br}, and A.A. Smirnov\thanks{%
e-mail: saa@dfn.if.usp.br}}
\date{\today }
\maketitle

\begin{abstract}
Various Green functions of the Dirac equation with a magnetic-solenoid field
(the superposition of the Aharonov-Bohm field and a collinear uniform
magnetic field) are constructed and studied. The problem is considered in $%
2+1$ and $3+1$ dimensions for the natural extension of the Dirac operator
(the extension obtained from the solenoid regularization). Representations
of the Green functions as proper time integrals are derived. The
nonrelativistic limit is considered. For the sake of completeness the Green
functions of the Klein-Gordon particles are constructed as well.
\end{abstract}

\section{Introduction}

In the present article we continue our previous study \cite
{BGT01,GGS03,GGS03a} of the Dirac equation with a magnetic-solenoid field,
constructing and studying various Green functions of this equation. We
recall that the magnetic-solenoid field is the collinear superposition of
the constant uniform magnetic field and the Aharonov-Bohm (AB) field. The AB
field is a field of an infinitely long and infinitesimally thin solenoid.
Recently the interest in such a field configuration has been renewed in
connection with planar physics problems, quantum Hall effect, and the\
Aharonov-Bohm effect in cyclotron and synchrotron radiations \cite
{N00,T00,C00,HO01,BGLT01,ESV02}.

In principle, the Green functions can be constructed whenever complete sets
of solutions of the Dirac equations are available. In this connection, one
ought to recall that solutions of the Dirac equation with the
magnetic-solenoid field in $2+1$ and $3+1$ dimensions were obtained in \cite
{BGT01}. The singularity of the AB field demands a special attention to the
correct definition of the Dirac operator. The need for self-adjoint
extensions in the case of the Dirac Hamiltonian with the pure AB field in $%
2+1$\ dimensions was recognized in \cite{GJ89,G89} where certain boundary
conditions at the origin were established. The regularized case and
peculiarities of the behavior of a spinning particle in the presence of the
magnetic string were considered in \cite{H90,H91}. The problem of the
self-adjoint extension of the Dirac operator with the magnetic-solenoid
field was studied in \cite{GGS03,GGS03a,FP01}. In $2+1$ dimensions, a
one-parametric family of \thinspace self-adjoint Dirac Hamiltonians
specified by the corresponding boundary conditions at the AB solenoid was
constructed, and the spectrum and eigenfunctions for each value of the
extension parameter were found. In $3+1$ dimensions, a two-parametric family
of the self-adjoint Dirac Hamiltonians was constructed on the condition that
the spin polarization\ is conserved. The corresponding spectrum and
eigenfunctions for each value of the extension parameters were found as
well. In \cite{GGS03,GGS03a} the procedure of solenoid regularization was
also considered. The procedure implies considering the finite solenoid and
then making its radius go to zero. This procedure specifies some particular
boundary conditions. The values of the extension parameters corresponding to
the solenoid regularization case were determined in $2+1$ and $3+1$
dimensions. Further, we call the corresponding extension the natural
extension. Nonrelativistic propagators for the spinless and spin-$1/2$
particle moving in the pure AB field were considered mainly in the relation
to the AB effect. The propagator of the spinless particle was found in \cite
{GS79,BI81,MM84} as a sum of partial propagators corresponding to
homotopically different paths in the covering space of the physical
background. The nonrelativistic propagator of the spin-$1/2$ particle in the
AB field for a particular value of the self-adjoint extension parameter was
discussed in \cite{Park95}. The relativistic scalar case for the AB field
was studied in \cite{GR91}. The propagators and the AB\ effect in general
gauge theories were considered in \cite{ST86,OSL88}. Recently, vacuum
polarization effects in the AB field have aroused great interest, see, for
example, \cite{BK99,BFKS00} and references therein.

In the present article, we construct and study the Green functions of the
Dirac particle in the magnetic-solenoid field in $2+1$ and $3+1$ dimensions.
The physical importance of the problem is stressed by the fact that the
knowledge of the Green functions in such a configuration allows one to study
quantum (and quantum field) effects in the magnetic-solenoid field on a
regular base. A technical specificity of the problem is related to the
necessity to take into account all the peculiarities related to the
self-adjoint extension problem of the Dirac operator in the background under
consideration. In Sec. 2 we consider the $\left( 2+1\right) $-dimensional
case in detail. Here, constructing the Green functions, we use the exact
solutions of the Dirac equation that are related to the specific values of
the extension parameter. These values correspond to the natural extension,
see above. The representations of the Green functions as proper time
integrals are derived. In addition, we calculate the nonrelativistic Green
functions as well. In Sec. 3 we extend the results to the $\left( 3+1\right)
$-dimensional case. In the Appendix, for the sake of completeness, we
present the Green functions of the relativistic scalar particle.

We note that the magnetic-solenoid field belongs to such a type of fields
that do not violate the vacuum stability. For such fields a unique stable
vacuum exists, and quantum field definitions of Green functions following
below hold true \cite{FGS91}. In particular, the causal propagator $%
S^{c}\left( x,x^{\prime }\right) $ and the anticausal propagator $S^{\bar{c}%
}\left( x,x^{\prime }\right) $ are defined by the expressions
\begin{eqnarray}
\ S^{c}\left( x,x^{\prime }\right) &=&i\left\langle 0\left| T\widehat{\psi }%
(x)\widehat{\overline{\psi }}(x^{\prime })\right| 0\right\rangle \,,
\label{w23} \\
S^{\bar{c}}\left( x,x^{\prime }\right) &=&i\left\langle 0\left| \widehat{%
\psi }(x)\widehat{\overline{\psi }}(x^{\prime })T\right| 0\right\rangle \,,
\label{w23k}
\end{eqnarray}
where $\widehat{\psi }(x)$ is the quantum spinor field in the Furry
representation, satisfying the Dirac equation with the magnetic-solenoid
field, $\left| 0\right\rangle $\ is the vacuum in this representation. The
symbol of the $T$-product acts on both sides: it orders the field operators
to its right side and antiorders them to its left. The functions $%
S^{c}\left( x,x^{\prime }\right) $, $S^{\bar{c}}\left( x,x^{\prime }\right) $
can be expressed via the functions $S^{\mp }(x,x^{\prime }),$%
\begin{eqnarray}
S^{c}\left( x,x^{\prime }\right) &=&\theta \left( \Delta x^{0}\right)
S^{-}\left( x,x^{\prime }\right) -\theta \left( -\Delta x^{0}\right)
S^{+}\left( x,x^{\prime }\right) ,\;\Delta x^{0}=x^{0}-x^{0\prime }\,,
\label{g1} \\
S^{\bar{c}}\left( x,x^{\prime }\right) &=&\theta \left( -\Delta x^{0}\right)
S^{-}\left( x,x^{\prime }\right) -\theta \left( \Delta x^{0}\right)
S^{+}\left( x,x^{\prime }\right) \,,  \label{g2}
\end{eqnarray}
and the latter can be calculated via a complete set $_{\pm }\psi _{a}\left(
x\right) $ of solutions of the Dirac equation with the magnetic-solenoid
field as
\begin{equation}
S^{\mp }(x,x^{\prime })=i\sum_{a}\,_{\pm }\psi _{a}\left( x\right) \,_{\pm }%
\overline{\psi }_{a}\left( x^{\prime }\right) \,.  \label{g4}
\end{equation}
The solutions with the subscript $(+)$ belong to the positive energy
spectrum, whereas the solutions with the subscript $(-)$ belong to the
negative energy spectrum. Via $a$ all possible quantum numbers are denoted.

The Dirac equation with the magnetic-solenoid field has the form
\begin{equation}
\left( \gamma ^{\nu }P_{\nu }-M\right) \psi \left( x\right) =0\,.  \label{e1}
\end{equation}
Here\ $P_{\nu }=i\partial _{\nu }-qA_{\nu }\left( x\right) $, $x=\left(
x^{\nu }\right) $, $q$ is an algebraic charge, for electrons $q=-e<0$, $M$
is the electron mass, and $A_{\nu }\left( x\right) $ are potentials of the
magnetic-solenoid field. In ($3+1$)-dimensional case $\nu =0,1,2,3$ and $%
\gamma ^{\nu }$ are the corresponding gamma-matrices. In ($2+1$)-dimensional
case $\nu =0,1,2$ and in what follows, we employ the letter $\Gamma $ to
denote the gamma-matrices. We use for these matrices the following
representation
\begin{equation*}
\Gamma ^{0}=\sigma ^{3},\Gamma ^{1}=i\sigma ^{2},\Gamma ^{2}=-i\sigma ^{1},
\end{equation*}
where $\sigma ^{i}$ are the Pauli matrices. In cylindric coordinates $\left(
\varphi ,r\right) $,$\;x^{1}=r\cos \varphi $, $x^{2}=r\sin \varphi $, the
potentials of the magnetic-solenoid field have the form
\begin{eqnarray}
&&A_{0}=0,\;eA_{1}=\left[ l_{0}+\mu +A\left( r\right) \right] \frac{\sin
\varphi }{r},\;eA_{2}=-\left[ l_{0}+\mu +A\left( r\right) \right] \frac{\cos
\varphi }{r}\,,  \notag \\
&&\,(A_{3}=0\;\mathrm{in\;}3+1{}),\;A\left( r\right) =eBr^{2}/2\,.
\label{abe2}
\end{eqnarray}
Here $B$\ is the magnitude of the uniform magnetic field, and the magnitude $%
B^{AB}$ of the AB field is given by the expression $B^{AB}=\Phi \delta
\left( x^{1}\right) \delta \left( x^{2}\right) $, where $\Phi $\ is the
AB-solenoid flux, $\left( l_{0}+\mu \right) =\Phi /\Phi _{0}$, $\Phi
_{0}=2\pi /e$. It is supposed that $l_{0}$ is integer and $0\leq \mu <1$.

The functions $S^{\mp }(x,x^{\prime })$ obey the Dirac equation (\ref{e1}),
whereas the causal and anticausal propagators obey the nonhomogeneous Dirac
equations:
\begin{equation*}
\left( \gamma ^{\nu }P_{\nu }-M\right) S^{c}\left( x,x^{\prime }\right)
=-\delta (x-x^{\prime })\,,\;\left( \gamma ^{\nu }P_{\nu }-M\right) S^{\bar{c%
}}\left( x,x^{\prime }\right) =\delta (x-x^{\prime })\,.
\end{equation*}
We note that the commutation function $S\left( x,x^{\prime }\right) $, the
advanced $S^{adv}\left( x,x^{\prime }\right) $ and the retarded $%
S^{ret}\left( x,x^{\prime }\right) $ Green functions can be expressed in
terms of $S^{c}\left( x,x^{\prime }\right) $, $S^{\bar{c}}\left( x,x^{\prime
}\right) $ as follows
\begin{eqnarray}
&&S\left( x,x^{\prime }\right) =S^{-}\left( x,x^{\prime }\right)
+S^{+}\left( x,x^{\prime }\right) =\mathrm{sgn}\left( \Delta x^{0}\right) %
\left[ S^{c}\left( x,x^{\prime }\right) -S^{\bar{c}}\left( x,x^{\prime
}\right) \right] \,,  \label{g7} \\
&&S^{adv}\left( x,x^{\prime }\right) =-\theta \left( -\Delta x^{0}\right)
S\left( x,x^{\prime }\right) ,\;S^{ret}\left( x,x^{\prime }\right) =\theta
\left( \Delta x^{0}\right) S\left( x,x^{\prime }\right) \,.  \label{g8a}
\end{eqnarray}

\section{2+1 dimensional case}

\subsection{Sets of exact solutions}

First we study the $\left( 2+1\right) $-dimensional case, for which, as
known \cite{GGS03,GGS03a}, the Dirac operator with the magnetic-solenoid
field in $2+1$ dimensions possesses a one-parameter family of self-adjoint
extensions. That provides a one-parameter family of boundary conditions at
the origin. Following \cite{GGS03,GGS03a}, we denote the extension parameter
as $\Theta $. Generally speaking, the AB symmetry is violated for the
spinning particle, which is therefore sensible to the solenoid flux sign. As
was demonstrated in \cite{GGS03,GGS03a}, the values $\Theta =\pm \pi /2$
correspond to the natural extension, $\Theta =-\pi /2$ if the flux is
positive and $\Theta =\pi /2$ if the flux is negative. Below we present a
set of solutions $_{\pm }\psi _{a}\left( x\right) $ of (\ref{e1}) which we
will use for Green function construction according to the formulas (\ref{g4}%
). We consider the problem separately for two values of the extension
parameter.

We start with the case $\Theta =-\pi /2$. The positive energy spectrum is
given by $_{+}\varepsilon $ and the negative energy spectrum is given by $%
_{-}\varepsilon \,,$
\begin{equation}
_{+}\varepsilon =-\,_{-}\varepsilon =\sqrt{M^{2}+\omega }\,.  \label{g9b}
\end{equation}
Both branches are determined by the spectrum of the quantity $\omega $ which
is defined below. The solutions $_{\pm }\psi _{a}\left( x\right) $ can be
expressed via the solutions $u\left( x\right) $ of the squared Dirac
equation. The latter solutions have the form
\begin{eqnarray}
&&_{\pm }u_{m,l,\sigma }\left( x\right) =e^{-i\,_{\pm }\varepsilon
x^{0}}u_{m,l,\sigma }\left( x_{\bot }\right) \,,  \notag \\
&&\,x_{\bot }=\left( x^{1},x^{2}\right) \,,\;m=0,1,\ldots \,,\;l=0,\pm
1,...\,,\;\sigma =\pm 1\,,  \label{g9}
\end{eqnarray}
where
\begin{eqnarray*}
&&u_{m,l,\sigma }\left( x_{\bot }\right) =\sqrt{\gamma }g_{l}\left( \varphi
\right) \phi _{m,l,\sigma }\left( r\right) \upsilon _{\sigma }\,,\;l\neq 0\,,
\\
&&u_{m,0,+1}\left( x_{\bot }\right) =\sqrt{\gamma }g_{0}\left( \varphi
\right) \phi _{m,0,+1}\left( r\right) \upsilon _{+1}\,, \\
&&u_{m,0,-1}\left( x_{\bot }\right) =\sqrt{\gamma }g_{0}\left( \varphi
\right) \phi _{m,-1}^{ir}\left( r\right) \upsilon _{-1}\,,\;\gamma =e\left|
B\right| \,,
\end{eqnarray*}
and
\begin{eqnarray*}
&&g_{l}(\varphi )=\frac{1}{\sqrt{2\pi }}\exp \left\{ i\varphi \left[ l-l_{0}-%
\frac{1}{2}\left( 1+\sigma ^{3}\right) \right] \right\} \,, \\
&&\upsilon _{+1}=\left(
\begin{array}{c}
1 \\
0
\end{array}
\right) ,\;\upsilon _{-1}=\left(
\begin{array}{c}
0 \\
1
\end{array}
\right) \,.
\end{eqnarray*}
The functions $\phi _{m,l,\sigma }\left( r\right) ,\;\phi _{m,-1}^{ir}(r)$
are expressed via the Laguerre functions $I_{m+\alpha ,m}(\rho )$ as
\begin{eqnarray}
&&\phi _{m,l,\sigma }(r)=I_{m+|\nu |,m}\left( \rho \right) \,,\;\phi
_{m,-1}^{ir}(r)=I_{m-\mu ,m}\left( \rho \right) \,,  \notag \\
&&\rho =\gamma r^{2}/2\,,\;\nu =\mu +l-\left( 1+\sigma \right) /2\,.
\label{abe20}
\end{eqnarray}
We recall that the Laguerre functions $I_{m+\alpha ,m}(\rho )$ are related
to the Laguerre polynomials $L_{m}^{\alpha }(x)$ (8.970, 8.972.1 \cite{GR94}%
) as
\begin{equation*}
I_{m+\alpha ,m}(x)=\sqrt{\frac{m!}{\Gamma \left( m+\alpha +1\right) }}%
e^{-x/2}x^{\alpha /2}L_{m}^{\alpha }(x)\,.
\end{equation*}
For the magnetic field $B>0$, the spectrum of $\omega $ corresponding to the
functions $u_{m,l,\sigma }\left( x_{\bot }\right) $ is
\begin{equation}
\omega =\left\{
\begin{array}{l}
2\gamma \left( m+l+\mu \right) ,\;{}l-\left( 1+\sigma \right) /2\geq 0 \\
2\gamma \left( m+\left( 1+\sigma \right) /2\right) ,\;{}l-\left( 1+\sigma
\right) /2<0
\end{array}
\right. ,  \label{abe21}
\end{equation}
except the functions $u_{m,0,-1}\left( x_{\bot }\right) $ for which the
spectrum of $\omega $ is
\begin{equation}
\omega =2\gamma m\,.  \label{abe22}
\end{equation}

Then the complete set $_{\pm }\psi _{a}$ with $a=\left( m,l\right) $ has the
form
\begin{equation}
\,_{\pm }\psi _{m,l}\left( x\right) =N\left( \Gamma P+M\right) \,_{\pm
}u_{m,l,-1}\left( x\right) \,.  \label{gx1}
\end{equation}
The latter form provides correct expressions both for $\omega \neq 0$ and $%
\omega =0,$ since the states with $\omega =0$ can only be expressed in terms
of the spinors with $\sigma =-1$ (we note that $\,_{+}\psi \equiv 0$ for $%
\omega =0$, nevertheless it is convenient to remain in (\ref{g9}) $u_{+}$
with $\,_{+}\varepsilon =M$). The normalization factor with respect to the
usual inner product $(\psi ,\psi ^{\prime })=\int \psi ^{\dagger }(x)\psi
^{\prime }(x)d\mathbf{x}$ reads
\begin{equation*}
N=\left\{
\begin{array}{c}
\left[ 2\left| _{\pm }\varepsilon \right| \left( \left| _{\pm }\varepsilon
\right| -M\right) \right] ^{-1/2}\,,\;\;\omega \neq 0\,, \\
\left[ 2M\right] ^{-1},\;\;\omega =0
\end{array}
\right. .
\end{equation*}
The quantum number{\Large \ }$l$ characterizes the angular momentum of the
particle, $m$ is the radial quantum number, see \cite{BGT01}.

For $B<0$ the spectrum of states differs nontrivially from the expressions
given by Eqs. (\ref{abe21}) and (\ref{abe22}). Here $\omega $ corresponding
to $u_{m,l,\sigma }(r)$ is
\begin{equation}
\omega =\left\{
\begin{array}{l}
2\gamma \left( m-l+1-\mu \right) ,\;{}l-\left( 1+\sigma \right) /2<0 \\
2\gamma \left( m+\left( 1-\sigma \right) /2\right) ,\;{}l-\left( 1+\sigma
\right) /2\geq 0
\end{array}
\right. ,  \label{abe23}
\end{equation}
except the functions $u_{m,0,-1}\left( x_{\bot }\right) $ for which the
spectrum of $\omega $ is
\begin{equation}
\omega =2\gamma \left( m+1-\mu \right) \,.  \label{abe24}
\end{equation}

Now we go to the case with the extension parameter $\Theta =\pi /2$. We
recall that one needs for self-adjoint extensions of the radial Dirac
Hamiltonian only in the subspace $l=0$ to which we refer as to the critical
subspace. Thus, the only solutions in the $l=0$ subspace must be subjected
to the one of asymptotic condition from a one-parametric family of boundary
conditions as $r\rightarrow 0$. By this reason for $\Theta =\pi /2$, the
solutions only differ from (\ref{g9}) in the subspace $l=0$,
\begin{eqnarray}
u_{m,0,+1}\left( x_{\bot }\right) &=&\sqrt{\gamma }g_{0}\left( \varphi
\right) \phi _{m,+1}^{ir}\left( r\right) \upsilon _{+1}\,,\;\;\phi
_{m,+1}^{ir}(r)=I_{m+\mu -1,m}\left( \rho \right) \,,  \notag \\
u_{m,0,-1}\left( x_{\bot }\right) &=&\sqrt{\gamma }g_{0}\left( \varphi
\right) \phi _{m,0,-1}\left( r\right) \upsilon _{-1}\,,  \label{g9a}
\end{eqnarray}
where the spectrum for $u_{m,0,+1}\left( x_{\bot }\right) $ is given as
\begin{eqnarray}
\omega &=&2\gamma \left( m+\mu \right) ,\;B>0\,,  \label{abe22a} \\
\omega &=&2\gamma m,\;B<0\,.  \label{abe24a}
\end{eqnarray}

\subsection{Construction of Green functions}

The main point in constructing the Green functions is the summations in the
representation (\ref{g4}). In the case under consideration, this summation
can be done with the help of special relations which can be established for
the solutions of the Dirac equation.

Let us start with the calculation of the Green functions for the extension
parameter $\Theta =-\pi /2$ and $B>0$. In this case, taking into account
that the eigenfunctions $u$ of the equation $\left[ \left( \Gamma P_{\bot
}\right) ^{2}+\omega \right] u=0$\ corresponding to any $\omega \neq 0$ obey
the equations
\begin{eqnarray}
&&\Gamma P_{\bot }\,_{\pm }u_{m_{+},l,-\sigma }\left( x\right) =-i\sqrt{%
\omega }\,_{\pm }u_{m_{-},l,\sigma }\left( x\right) ,\;l\leq 0\,,\;P_{\bot
}=\left( 0,P_{1},P_{2}\right) \,,  \notag \\
&&\Gamma P_{\bot }\,_{\pm }u_{m,l,-\sigma }\left( x\right) =i\sqrt{\omega }%
\,_{\pm }u_{m,l,\sigma }\left( x\right) ,\;l\geq 1,\;m_{\pm }=m+\left( 1\pm
\sigma \right) /2\,,  \label{g10}
\end{eqnarray}
and the explicit form of the solutions $_{\pm }\psi _{m,l}$, one can verify
that for $\left| \varepsilon \right| \neq M$ the following relations hold
true,
\begin{eqnarray}
\,_{\pm }\psi _{m,l}\left( x\right) \,_{\pm }\overline{\psi }_{m,l}\left(
x^{\prime }\right) &=&\left( \Gamma P+M\right) \frac{1}{2\,_{\pm
}\varepsilon }e^{-i\,_{\pm }\varepsilon \Delta x^{0}}\sum_{\sigma =\pm
1}\phi _{m_{-},l,\sigma }\left( x_{\bot },x_{\bot }^{\prime }\right) \Xi
_{\sigma }\,,\,\;l\leq 0\,,  \notag \\
\,_{\pm }\psi _{m,l}\left( x\right) \,_{\pm }\overline{\psi }_{m,l}\left(
x^{\prime }\right) &=&\left( \Gamma P+M\right) \frac{1}{2\,_{\pm
}\varepsilon }e^{-i\,_{\pm }\varepsilon \Delta x^{0}}\sum_{\sigma =\pm
1}\phi _{m,l,\sigma }\left( x_{\bot },x_{\bot }^{\prime }\right) \Xi
_{\sigma }\,,\,\;l\geq 1\,,  \label{g12}
\end{eqnarray}
where
\begin{eqnarray}
&&\phi _{m,l,\sigma }\left( x_{\bot },x_{\bot }^{\prime }\right) =\frac{%
\gamma }{2\pi }e^{i\left[ l-l_{0}-\left( 1+\sigma \right) /2\right] \Delta
\varphi }I_{m+\alpha ,m}\left( \rho \right) I_{m+\alpha ,m}\left( \rho
^{\prime }\right) \,,  \label{g13} \\
&&\Delta \varphi =\varphi -\varphi ^{\prime },\;\;\alpha =\left\{
\begin{array}{c}
\mu +l-\left( 1+\sigma \right) /2,\;l\geq 1 \\
-\left[ \mu +l-\left( 1+\sigma \right) /2\right] ,\;l\leq 0
\end{array}
\right. ,\;\;\Xi _{\pm 1}=\left( 1\pm \sigma ^{3}\right) /2\,.  \notag
\end{eqnarray}
The above relations and Eqs. (\ref{g1}), (\ref{g4}) allow us to represent
the causal Green function in the following form
\begin{eqnarray}
&&S^{c}\left( x,x^{\prime }\right) =\left( \Gamma P+M\right) \Delta
^{c}\left( x,x^{\prime }\right) \,,  \notag \\
&&\Delta ^{c}\left( x,x^{\prime }\right) =i\sum_{m,l,\sigma }\left[ \theta
\left( \Delta x^{0}\right) \frac{e^{-i\,_{+}\varepsilon \Delta x^{0}}}{%
2\,_{+}\varepsilon }-\theta \left( -\Delta x^{0}\right) \frac{%
e^{-i\,_{-}\varepsilon \Delta x^{0}}}{2\,_{-}\varepsilon }\right] \phi
_{m,l,\sigma }\left( x_{\bot },x_{\bot }^{\prime }\right) \Xi _{\sigma }\,.
\label{g21}
\end{eqnarray}
Then we can use the representations
\begin{eqnarray}
&&\theta \left( \Delta x^{0}\right) \frac{e^{-i\,_{+}\varepsilon \Delta
x^{0}}}{2\,_{+}\varepsilon }-\theta \left( -\Delta x^{0}\right) \frac{%
e^{-i\,_{-}\varepsilon \Delta x^{0}}}{2\,_{-}\varepsilon }=\frac{1}{2\pi i}%
\int_{-\infty }^{\infty }\frac{e^{-ip_{0}\Delta x^{0}}}{\varepsilon
^{2}-p_{0}^{2}-i\epsilon }dp_{0}\,,  \label{g15} \\
&&\frac{1}{\varepsilon ^{2}-p_{0}^{2}-i\epsilon }=i\int_{0}^{\infty
}e^{-i\left( \varepsilon ^{2}-p_{0}^{2}\right) s}ds\,,  \label{g16}
\end{eqnarray}
in Eq. (\ref{g21}). Integrating over $p_{0},$ we obtain finally
\begin{eqnarray}
\Delta ^{c}\left( x,x^{\prime }\right) &=&\int_{0}^{\infty }f\left(
x,x^{\prime },s\right) ds\,,  \notag \\
f\left( x,x^{\prime },s\right) &=&\frac{1}{2\left( \pi s\right) ^{1/2}}e^{%
\frac{-i\Delta x_{0}^{2}}{4s}}e^{i\pi /4}e^{-iM^{2}s}i\sum_{m,l,\sigma
}e^{-i\omega s}\phi _{m,l,\sigma }\left( x_{\bot },x_{\bot }^{\prime
}\right) \Xi _{\sigma }\,.  \label{g22}
\end{eqnarray}
The path of the integration over $s$ is deformed so that it goes slightly
below the singular points $s_{k}=k\pi /\gamma $, $k=1,2,...$ .

Using \ (\ref{g4}), (\ref{g12}), and the representation
\begin{eqnarray}
&&\left. -\right. \theta \left( -\Delta x^{0}\right) \frac{%
e^{-i\,_{+}\varepsilon \Delta x^{0}}}{2\,_{+}\varepsilon }+\theta \left(
\Delta x^{0}\right) \frac{e^{-i\,_{-}\varepsilon \Delta x^{0}}}{%
2\,_{-}\varepsilon }=\frac{1}{2\pi i}\int_{-\infty }^{\infty }\frac{%
e^{-ip_{0}\Delta x^{0}}}{\varepsilon ^{2}-p_{0}^{2}+i\epsilon }dp_{0}\,,
\notag \\
&&\frac{1}{\varepsilon ^{2}-p_{0}^{2}+i\epsilon }=i\int_{-0}^{-\infty
}e^{-i\left( \varepsilon ^{2}-p_{0}^{2}\right) s}ds\,.  \label{g16a}
\end{eqnarray}
instead of (\ref{g15}), (\ref{g16}) we obtain from (\ref{g2}),
\begin{equation}
S^{\bar{c}}\left( x,x^{\prime }\right) =\left( \Gamma P+M\right) \Delta ^{%
\bar{c}}\left( x,x^{\prime }\right) ,\;\;\Delta ^{\bar{c}}\left( x,x^{\prime
}\right) =\int_{-0}^{-\infty }f\left( x,x^{\prime },s\right) ds\,,
\label{g23}
\end{equation}
where $f\left( x,x^{\prime },s\right) $ is given by Eq. (\ref{g22}). The
negative values for $s$ are defined as $s=\left| s\right| e^{-i\pi }$, and
the path of integration over $s$ is deformed so that it goes slightly below
the singular points $-s_{k}$.

We now consider the summations in (\ref{g22}). Applying the formula (8.976
(1) \cite{GR94}) we can sum over $m$ to get
\begin{eqnarray}
&&\sum_{m=0}^{\infty }e^{-i2m\gamma s}I_{m+\alpha ,m}\left( \rho \right)
I_{m+\alpha ,m}\left( \rho ^{\prime }\right) =\exp \left\{ \frac{i}{2}\left(
\rho +\rho ^{\prime }\right) \cot \left( \gamma s\right) \right\}  \notag \\
&&\times \frac{e^{i\alpha \gamma s}e^{i\gamma s}}{2i\sin \left( \gamma
s\right) }e^{-\frac{i\pi \alpha }{2}}J_{\alpha }\left( z\right) ,\;z\left.
=\right. \sqrt{\rho \rho ^{\prime }}/\sin \left( \gamma s\right) \,,
\label{g24}
\end{eqnarray}
where $J_{\alpha }\left( z\right) $ are the Bessel functions (8.402 \cite
{GR94}), and for negative $s$ we take $\arg s=-\pi +0$.

Similar results can be obtained for the case $B<0$. Here one should use the
solutions corresponding to the spectrum of $\omega $ (\ref{abe23}), (\ref
{abe24}). Then these results can be united to obtain expressions which hold
true for any sign of $B$,
\begin{eqnarray}
&&f\left( x,x^{\prime },s\right) =\sum_{l=-\infty }^{\infty }f_{l}\left(
x,x^{\prime },s\right) ,\;\;f_{l}\left( x,x^{\prime },s\right) =A\left(
s\right) \sum_{\sigma =\pm 1}\Phi _{l,\sigma }\left( s\right) e^{-i\sigma
eBs}\Xi _{\sigma }\,,  \notag \\
&&A\left( s\right) =\frac{eB}{8\pi ^{3/2}s^{1/2}\sin \left( eBs\right) }\exp
\left\{ \frac{i\pi }{4}-iM^{2}s-il_{0}\Delta \varphi \right\}  \notag \\
&&\times \exp \left\{ -\frac{i\left( \Delta x_{0}\right) ^{2}}{4s}+\frac{ieB%
}{4}\left( r^{2}+r^{\prime 2}\right) \cot \left( eBs\right) \right\} \,,
\label{g250} \\
&&\Phi _{l,\sigma }\left( s\right) =e^{il_{\sigma }\Delta \varphi
}e^{-i\left( l_{\sigma }+\mu \right) eBs}e^{-\frac{i\pi \left| l_{\sigma
}+\mu \right| }{2}}J_{\left| l_{\sigma }+\mu \right| }\left( z\right)
,\;l_{\sigma }=l-\left( 1+\sigma \right) /2,\;l\neq 0\,,  \notag \\
&&\Phi _{0,+1}\left( s\right) =e^{-i\Delta \varphi }e^{i\left( 1-\mu \right)
eBs}e^{-\frac{i\pi \left( 1-\mu \right) }{2}}J_{1-\mu }\left( z\right)
,\;\Phi _{0,-1}\left( s\right) =e^{-i\mu eBs}e^{\frac{i\pi \mu }{2}}J_{-\mu
}\left( z\right) \,.  \label{g25a}
\end{eqnarray}

Now we consider the summation over $l$. One can see that the following
relations hold true
\begin{eqnarray*}
\sum_{l=1}^{\infty }\Phi _{l,-1}\left( s\right) &=&\sum_{l=1}^{\infty }\Phi
_{l+1,+1}\left( s\right) =e^{-i\mu eBs}Y\left( z,\Delta \varphi -eBs,\mu
\right) \,, \\
\sum_{l=-1}^{-\infty }\Phi _{l,-1}\left( s\right) &=&\sum_{l=-1}^{-\infty
}\Phi _{l+1,+1}\left( s\right) =e^{-i\mu eBs}Y\left( z,-\Delta \varphi
+eBs,-\mu \right) \,,
\end{eqnarray*}
where
\begin{equation}
Y\left( z,\eta ,\mu \right) =a_{1}\left( z\right) +\widetilde{Y}\left(
z,\eta ,\mu \right) ,\;\widetilde{Y}\left( z,\eta ,\mu \right)
=\sum_{l=2}^{\infty }a_{l}\left( z\right) ,\;a_{l}\left( z\right) =e^{i\eta
l}\left( -i\right) ^{l+\mu }J_{l+\mu }\left( z\right) \,.  \label{y-intro}
\end{equation}
The evaluation of the sum in (\ref{y-intro}) can be done in a similar way to
what was done in \cite{AB59}. There exist all $\partial _{z}a_{l}\left(
z\right) $ on the half-line, $0<z<\infty $, and the relation (8.471 (2) \cite
{GR94}), $\partial _{z}J_{\nu }\left( z\right) =\left[ J_{\nu -1}\left(
z\right) -J_{\nu +1}\left( z\right) \right] /2$, can be used. The series $%
\widetilde{Y}\left( z,\eta ,\mu \right) $ converges and the series of
derivatives $\sum_{l=2}^{\infty }\partial _{z}a_{l}\left( z\right) $
converges uniformly in $\left( 0,\infty \right) $. It is sufficient
condition to write down $\partial _{z}\widetilde{Y}\left( z,\eta ,\mu
\right) =\sum_{l=2}^{\infty }\partial _{z}a_{l}\left( z\right) $. Thus, one
arrives to a differential equation with respect to $Y\left( z,\eta ,\mu
\right) $,
\begin{equation}
\frac{d}{dz}Y\left( z,\eta ,\mu \right) =-Y\left( z,\eta ,\mu \right) i\cos
\eta +\frac{1}{2}\left( -i\right) ^{\mu }\left[ -ie^{i\eta }J_{\mu }\left(
z\right) +J_{1+\mu }\left( z\right) \right] \,.  \label{diff-eq}
\end{equation}
that is true on the half-line, $0<z<\infty $. The solution of (\ref{diff-eq}%
) reads
\begin{equation}
Y\left( z,\eta ,\mu \right) =\frac{1}{2}\left( -i\right) ^{\mu
}\int_{0}^{z}e^{i\left( y-z\right) \cos \eta }\left[ -ie^{i\eta }J_{\mu
}\left( y\right) +J_{1+\mu }\left( y\right) \right] dy\,.  \label{fsum}
\end{equation}
This is also valid for $Y\left( z,-\eta ,-\mu \right) $.

It is useful to introduce the following function
\begin{equation*}
f_{nc}\left( x,x^{\prime },s\right) =\sum_{l\neq 0}f_{l}\left( x,x^{\prime
},s\right) \,.
\end{equation*}
It defines the part of the Green functions that is the same for all
extensions. With the help of the function $Y\left( z,\eta ,\mu \right) $ (%
\ref{y-intro}), (\ref{fsum}) one can write
\begin{eqnarray}
&&f_{nc}\left( x,x^{\prime },s\right) =A\left( s\right) e^{-i\mu
eBs}e^{-ieBs\sigma ^{3}}\left\{ Y\left( z,\Delta \varphi -eBs,\mu \right)
+Y\left( z,-\Delta \varphi +eBs,-\mu \right) \right.  \notag \\
&&\left. +\left[ e^{-\frac{i\pi \mu }{2}}J_{\mu }\left( z\right)
-e^{-i\left( \Delta \varphi -eBs\right) }e^{-\frac{i\pi \left( 1-\mu \right)
}{2}}J_{1-\mu }\left( z\right) \right] \Xi _{+1}\right\} \,.  \label{g25}
\end{eqnarray}
The function $f_{0}\left( x,x^{\prime },s\right) $ is specific for each
extension. It is reasonable to mark it with a superscript that assumes the
values of the extension parameter. Thus, for $\Theta =-\pi /2$,
\begin{equation}
f_{0}^{\left( -\pi /2\right) }\left( x,x^{\prime },s\right) =A\left(
s\right) e^{-i\mu eBs}\left[ e^{-i\Delta \varphi }e^{-\frac{i\pi \left(
1-\mu \right) }{2}}J_{1-\mu }\left( z\right) \Xi _{+1}+e^{-ieBs\sigma
^{3}}e^{\frac{i\pi \mu }{2}}J_{-\mu }\left( z\right) \Xi _{-1}\right] \,.
\label{g26}
\end{equation}
Accordingly, the function $f\left( x,x^{\prime },s\right) $ acquires the
same superscript,
\begin{equation}
f^{\left( -\pi /2\right) }\left( x,x^{\prime },s\right) =f_{nc}\left(
x,x^{\prime },s\right) +f_{0}^{\left( -\pi /2\right) }\left( x,x^{\prime
},s\right) \,.  \label{g27}
\end{equation}

For the extension parameter $\Theta =\pi /2$, one obtains
\begin{eqnarray}
&&f_{0}^{\left( \pi /2\right) }\left( x,x^{\prime },s\right) =A\left(
s\right) e^{-i\mu eBs}\left[ e^{-i\Delta \varphi }e^{-\frac{i\pi \left( \mu
-1\right) }{2}}J_{\mu -1}\left( z\right) \Xi _{+1}+e^{-ieBs\sigma ^{3}}e^{-%
\frac{i\pi \mu }{2}}J_{\mu }\left( z\right) \Xi _{-1}\right] \,,  \notag \\
&&f^{\left( \pi /2\right) }\left( x,x^{\prime },s\right) =f_{nc}\left(
x,x^{\prime },s\right) +f_{0}^{\left( \pi /2\right) }\left( x,x^{\prime
},s\right) \,.  \label{g29}
\end{eqnarray}

Besides, one can consider particles with ''spin down'' polarization in $2+1$
dimensions. The corresponding wave functions $\psi ^{(-1)}\left( x\right) $
can be presented as
\begin{equation*}
\psi ^{(-1)}\left( x\right) =\sigma ^{1}\left( \Gamma P-M\right) u\left(
x\right) \,,
\end{equation*}
where $u\left( x\right) $ are solutions (\ref{g9}) of the squared Dirac
equation. The propagator related to such particles\ can be expressed in
terms of the function $\Delta ^{c}\left( x,x^{\prime }\right) $ (\ref{g21}),
\begin{equation*}
S_{\left( -1\right) }^{c}\left( x,x^{\prime }\right) =-\sigma ^{1}\left(
\Gamma P-M\right) \Delta ^{c}\left( x,x^{\prime }\right) \sigma ^{1}\,.
\end{equation*}

At this point we ought to make some remarks.

One can see that there exists a simple relation between scalar Green
functions and Green functions of the squared Dirac equation (for the above
considered extensions). Consider this relation in the example of causal
Green functions. First of all, we note that the Klein-Gordon equation
differs from the squared Dirac equation by the Zeeman interaction term. Then
we can see (remembering the origin of the quantum number $l$ for both
spinning and spinless particles) that the scalar propagator can be derived
from $\Delta ^{c}\left( x,x^{\prime }\right) $ by only retaining the terms
with $\sigma =-1$ only. The term $eB\sigma ^{3},$ which is responsible for
the Zeeman interaction with the uniform magnetic field, has to be removed.
The Zeeman interaction with the solenoid flux, influencing the terms with $%
l=0$, depends on the flux sign and can be repulsive or attractive. The
repulsive contact interaction case is physically equivalent to the spinless
case, since in both cases the corresponding wave functions vanish at the
origin. The necessary boundary condition is realized for the extension
parameter $\Theta =\pi /2$. Thus, one can obtain the scalar Green functions
using the coefficients of $\Xi _{-1}$ in $f_{l}\left( x,x^{\prime },s\right)
$ (\ref{g250}), (\ref{g25a}) and $f_{0}^{\left( \pi /2\right) }\left(
x,x^{\prime },s\right) $ (\ref{g29}). By following such prescriptions, one
arrives at the expression (\ref{a2}) obtained by direct calculation.

In the spinless case there is no physically preferred orientation of the
plane $x^{1}x^{2}$. Therefore, the solenoid flux direction does not matter,
i.e., the AB symmetry, $l_{0}\rightarrow l_{0}+1$, is conserved. The
direction of the uniform magnetic field does not matter as well. This can be
observed from the explicit form of the Green functions (\ref{a2}) where the
change $B\rightarrow -B$\ is equivalent to the choice of the opposite
orientation of the plane, $l\rightarrow -l$, $\Delta \varphi \rightarrow
-\Delta \varphi $, $\Phi \rightarrow -\Phi $. In the spinning case the given
spin direction breaks the symmetry related to the plane orientation. The
Zeeman interaction of the spin with the background violates the AB symmetry
as well as the symmetry with respect to the change $B\rightarrow -B$.

As is known, influence of the solenoid flux on the particle is observed only
when the flux is not equal to an integral number of quanta ($\mu \neq 0$).
In this connection it is instructive to consider the Green functions for the
particular case $\mu =0$. We note that the part $f_{nc}\left( x,x^{\prime
},s\right) $ (\ref{g25}) of the function $f\left( x,x^{\prime },s\right) $\
is regular everywhere, while the part $f_{0}\left( x,x^{\prime },s\right) $
is singular at the origin. Thus, taking the limit $\mu \rightarrow 0$ in (%
\ref{g25}) and using the relation $J_{1}\left( y\right) =-J_{0}^{\prime
}\left( y\right) $ we get,
\begin{equation*}
f_{nc}\left( x,x^{\prime },s\right) =A\left( s\right) e^{-ieBs\sigma
^{3}}\left\{ e^{-iz\cos \left( \Delta \varphi -eBs\right) }-J_{0}\left(
z\right) +\left[ J_{0}\left( z\right) +ie^{-i\left( \Delta \varphi
-eBs\right) }J_{1}\left( z\right) \right] \Xi _{+1}\right\} \,.
\end{equation*}
The corresponding expression for $f_{0}\left( x,x^{\prime },s\right) $ can
be obtained in the following way. We restrict the range of $z$\ to $0<\delta
<z<\infty $, where $\delta \ll 1$. Then we take the limit $\mu \rightarrow 0$
and use the continuity of the Bessel functions with respect to its index. At
the end we construct the analytic continuation of the obtained expressions
over the interval $\left( 0,\delta \right) $. Thus, starting from either (%
\ref{g26}) or (\ref{g29}) we get,
\begin{equation*}
f_{0}\left( x,x^{\prime },s\right) =A\left( s\right) \left[ -ie^{-i\Delta
\varphi }J_{1}\left( z\right) \Xi _{+1}+e^{ieBs}J_{0}\left( z\right) \Xi
_{-1}\right] \,,
\end{equation*}
where the superscript is no longer necessary. Thus, the explicit form of $%
f\left( x,x^{\prime },s\right) $ is
\begin{eqnarray}
&&f\left( x,x^{\prime },s\right) =\frac{eB}{8\pi ^{3/2}s^{1/2}\sin \left(
eBs\right) }\exp \left\{ \frac{i\pi }{4}-\frac{i\left( \Delta x_{0}\right)
^{2}}{4s}-iM^{2}s-ieBs\sigma ^{3}\right\}  \notag \\
&&\times \exp \left\{ -il_{0}\Delta \varphi +\frac{ieB}{4}\left(
r^{2}+r^{\prime 2}\right) \cot \left( eBs\right) -\frac{ieBrr^{\prime }\cos
\left( \Delta \varphi -eBs\right) }{2\sin \left( eBs\right) }\right\} \,.
\label{gg1}
\end{eqnarray}
Making a transformation to Cartesian coordinates in (\ref{gg1}) and setting $%
l_{0}=0,$ one can obtain the known result of the uniform magnetic field, see
for example \cite{GGG98}.

\subsection{Nonrelativistic case}

Consideration of the Green functions in the background under question in the
nonrelativistic case is important for various physical applications. Below
we study this case in detail. The solutions of the Schr\"{o}dinger equation
for ''spin up'' particles ($+$) and antiparticles ($-$) in the case $\Theta
=-\pi /2$ read,
\begin{eqnarray}
\,_{+}\phi _{m,l}\left( x\right) &=&e^{-iEx^{0}}\sqrt{\frac{\gamma }{2\pi }}%
e^{i\left( l-l_{0}-1\right) \varphi }\phi _{m,l,+1}\left( r\right) ,\;E=%
\frac{\omega _{m,l,\sigma }}{2M}\,,  \label{g40} \\
\,_{-}\phi _{m,l}\left( x\right) &=&e^{-iEx^{0}}\sqrt{\frac{\gamma }{2\pi }}%
e^{-i\left( l-l_{0}\right) \varphi }\phi _{m,l,-1}\left( r\right) ,\;l\neq
0\,,  \notag \\
\,_{-}\phi _{m,0}\left( x\right) &=&e^{-iEx^{0}}\sqrt{\frac{\gamma }{2\pi }}%
e^{il_{0}\varphi }\phi _{m,-1}^{ir}\left( r\right) \,,  \label{g41}
\end{eqnarray}
where the values $\omega _{m,l,\sigma }$\ are defined by $m,l,\sigma $ with
the help of formulas (\ref{abe21}), (\ref{abe22}) for $B>0$, and (\ref{abe23}%
), (\ref{abe24}) for $B<0$. The solutions $\,_{+}\phi _{m,l}\left( x\right) $
($\,_{-}\phi _{m,l}\left( x\right) $ ) for\ the ''spin down'' case can be
obtained from the solutions $\,_{-}\phi _{m,l}\left( x\right) $ ($\,_{+}\phi
_{m,l}\left( x\right) $ ) for\ the ''spin-up'' case with the change $\varphi
\rightarrow -\varphi $ in (\ref{g40}), (\ref{g41}).

The retarded Green functions for particles and antiparticles are defined as
\begin{eqnarray}
&&S^{ret,\left( \pm \right) }\left( x,x^{\prime }\right) =\theta \left(
\Delta x^{0}\right) \sum_{l}S_{l}^{\left( \pm \right) }\left( x,x^{\prime
}\right) ,\;S_{l}^{\left( \pm \right) }\left( x,x^{\prime }\right)
=i\sum_{m}\,_{\pm }\phi _{m,l}\left( x\right) \,_{\pm }\phi _{m,l}^{\ast
}\left( x^{\prime }\right) \,,  \notag \\
&&S_{nc}^{\left( \pm \right) }\left( x,x^{\prime }\right) =\sum_{l\neq
0}S_{l}^{\left( \pm \right) }\left( x,x^{\prime }\right) \,,  \label{g42}
\end{eqnarray}
where the part $S_{nc}^{\left( \pm \right) }\left( x,x^{\prime }\right) $ is
the same for all extensions, whereas $S_{0}^{\left( \pm \right) }\left(
x,x^{\prime }\right) $ is specific for each extension. Carrying out the
summations in (\ref{g42}) one obtains
\begin{eqnarray}
&&S_{l}^{\left( \pm \right) }\left( x,x^{\prime }\right) =A_{nr}\left(
x,x^{\prime }\right) e^{\mp i\gamma \tau }e^{\pm i\left( l_{\pm
}-l_{0}\right) \Delta \varphi }e^{-i\left| l_{\pm }+\mu \right| \gamma \tau
}e^{-\frac{i\pi \left| l_{\pm }+\mu \right| }{2}}J_{\left| l_{\pm }+\mu
\right| }\left( z_{nr}\right) \,,  \notag \\
&&A_{nr}\left( x,x^{\prime }\right) =\frac{\gamma }{4\pi \sin \left( \gamma
\tau \right) }\exp \left[ \frac{i}{2}\left( \rho +\rho ^{\prime }\right)
\cot \left( \gamma \tau \right) \right] \,,  \label{g43} \\
&&S_{nc}^{\left( +\right) }\left( x,x^{\prime }\right) =A_{nr}\left(
x,x^{\prime }\right) e^{-il_{0}\Delta \varphi }e^{-i\left( 1+\mu \right)
eB\tau }\left\{ e^{-\frac{i\pi \mu }{2}}J_{\mu }\left( z_{nr}\right)
-e^{-i\Delta \varphi }e^{ieB\tau }e^{-\frac{i\pi \left( 1-\mu \right) }{2}%
}J_{1-\mu }\left( z_{nr}\right) \right.  \notag \\
&&\left. +Y\left( z_{nr},\Delta \varphi -eB\tau ,\mu \right) +Y\left(
z_{nr},-\Delta \varphi +eB\tau ,-\mu \right) \right\} \,,  \label{g44} \\
&&S_{nc}^{\left( -\right) }\left( x,x^{\prime }\right) =A_{nr}\left(
x,x^{\prime }\right) e^{il_{0}\Delta \varphi }e^{i\left( 1-\mu \right)
eB\tau }  \notag \\
&&\times \left\{ Y\left( z_{nr},-\Delta \varphi -eB\tau ,\mu \right)
+Y\left( z_{nr},\Delta \varphi +eB\tau ,-\mu \right) \right\} ,  \label{g45}
\\
&&z_{nr}=\sqrt{\rho \rho ^{\prime }}/\sin \left( \gamma \tau \right) ,\;\tau
=\Delta x^{0}/2M\,,\;l_{\pm }\left. =\right. l-\left( 1\pm 1\right)
/2,\;l\left. \neq \right. 0\,,  \notag
\end{eqnarray}
whereas for $l=0$,
\begin{eqnarray}
S_{0}^{\left( +\right) \left( \mp \pi /2\right) }\left( x,x^{\prime }\right)
&=&A_{nr}\left( x,x^{\prime }\right) e^{-i\left( l_{0}+1\right) \Delta
\varphi }e^{-i\mu eB\tau }e^{\mp \frac{i\pi \left( 1-\mu \right) }{2}}J_{\pm
\left( 1-\mu \right) }\left( z_{nr}\right) \,,  \label{g46} \\
S_{0}^{\left( -\right) \left( \mp \pi /2\right) }\left( x,x^{\prime }\right)
&=&A_{nr}\left( x,x^{\prime }\right) e^{il_{0}\Delta \varphi }e^{i\left(
1-\mu \right) eB\tau }e^{\pm \frac{i\pi \mu }{2}}J_{\mp \mu }\left(
z_{nr}\right) \,.  \label{g47}
\end{eqnarray}
The Green function in the ''spin down'' case can be obtained with the change
$\Delta \varphi \rightarrow -\Delta \varphi $ in (\ref{g43})-(\ref{g47}) and
with the change $S^{\left( \pm \right) }$ by $S^{\left( \mp \right) }$ in
all the functions $S\left( x,x^{\prime }\right) $ in (\ref{g43})-(\ref{g47}%
). Thus, one can see that the Green functions for the nonrelativistic
particle is irregular at $r=0$ when the contact interaction is attractive.

We note that for the limiting case $B=0$ (the uniform magnetic field is
absent) $S_{l}^{\left( +\right) \left( -\pi /2\right) }\left( x,x^{\prime
}\right) $ coincide with the known expression for the spinless particle \cite
{GS79,BI81,MM84}, which is natural in the case of a repulsive contact
interaction. While $S_{l}^{\left( +\right) \left( \pi /2\right) }\left(
x,x^{\prime }\right) $ for $B=0$ coincide with the corresponding expressions
obtained in the paper \cite{Park95}.

\section{3+1 dimensional case}

To obtain the Green functions in $3+1$ dimensions we use the orthonormalized
solutions $_{\pm }\Psi _{p_{3},m,l,\sigma }\left( x\right) $ of the Dirac
equation found in \cite{GGS03,GGS03a}. The quantum numbers $m$, $l$ have the
same meaning as in the $\left( 2+1\right) $-dimensional case, $p_{3}$\ is
the $x^{3}$-component of the momentum, and $\sigma $ is the spin quantum
number.{\LARGE \ }The positive energy spectrum is given by $_{+}\varepsilon $
and the negative energy spectrum is given by $_{-}\varepsilon \,.$ They both
are expressed via the quantity $\omega $ as
\begin{equation}
_{+}\varepsilon =-\,_{-}\varepsilon =\sqrt{M^{2}+p_{3}^{2}+\omega }\,.
\label{g30a}
\end{equation}
The spectra of $\omega $\ are given in (\ref{abe21}), (\ref{abe22}) for $B>0$%
, and in (\ref{abe23}), (\ref{abe24}) for $B<0$. For $\omega \neq 0$, one
can present the solutions $_{\pm }\Psi _{p_{3},m,l,\sigma }$ in the
following form,
\begin{eqnarray}
\, &&_{\pm }\Psi _{p_{3},m,l,\sigma }\left( x\right) =N\left( \gamma ^{\nu
}P_{\nu }+M\right) \,_{\pm }U_{p_{3},m,l,\sigma }\left( x\right) \,,  \notag
\\
&&\,_{\pm }U_{p_{3},m,l,\sigma }\left( x\right) =\frac{1}{\sqrt{2\pi }}%
e^{-i\,_{\pm }\varepsilon x^{0}-ip_{3}x^{3}}U_{m,l,\sigma }\left( x_{\bot
}\right) \,,  \notag \\
&&U_{m,l,\sigma }\left( x_{\bot }\right) =\left(
\begin{array}{c}
u_{m,l,\sigma }\left( x_{\bot }\right) \\
\sigma ^{3}u_{m,l,\sigma }\left( x_{\bot }\right)
\end{array}
\right) \,,\;N=\left[ 2\left| _{\pm }\varepsilon \right| \left( \left| _{\pm
}\varepsilon \right| +p_{3}\right) \right] ^{-1/2}\,,  \label{g30}
\end{eqnarray}
whereas for $\omega =0$,
\begin{equation*}
\,_{\pm }\Psi _{p_{3},0,l,-\xi }\left( x\right) =N\left( \gamma ^{\nu
}P_{\nu }+M\right) \,_{\pm }U_{p_{3},0,l,-\xi }\left( x\right) ,\;\xi =\text{%
\textrm{sgn}}\left( B\right) \,,
\end{equation*}
where $u_{m,l,\sigma }\left( x_{\bot }\right) $ are the two-spinors defined
in (\ref{g9}).

We are going to construct the Green functions using the solutions that
correspond to the natural extensions of the Dirac operator, i.e., for the
extension parameters chosen as $\Theta _{+1}=\Theta _{-1}=\Theta $, and $%
\Theta =\pm \pi /2$. First we consider the case $\Theta =-\pi /2$, and $B>0$%
. We note that for $\omega \neq 0$,
\begin{eqnarray}
&&\gamma ^{\nu }P_{\bot \nu }U_{m,l,-\sigma }=i\sqrt{\omega }U_{m,l,\sigma
}\,,\;l\geq 1\,,  \notag \\
&&\gamma ^{\nu }P_{\bot \nu }U_{m_{+},l,-\sigma }=-i\sqrt{\omega }%
U_{m_{-},l,\sigma }\,,\;l\leq 0\,,  \label{g32}
\end{eqnarray}
where $P_{\bot }=\left( 0,P_{1},P_{2},0\right) $. The summations in (\ref{g4}%
) can be done similarly to the ($2+1)$-dimensional case by the help of some
important relations derived by us for the solutions (\ref{g30}). Namely, for
the states with a given $\omega \neq 0$, the following relations hold true
\begin{eqnarray}
&&\sum_{\sigma =\pm 1}\,_{\pm }\Psi _{p_{3},m,l,\sigma }\left( x\right)
\,_{\pm }\overline{\Psi }_{p_{3},m,l,\sigma }\left( x^{\prime }\right)
\notag \\
&=&\sum_{\sigma =\pm 1}\frac{1}{2\,_{\pm }\varepsilon }\left( \gamma ^{\nu
}P_{\nu }+M\right) \frac{1}{2}\left( 1+\sigma \Sigma ^{3}\right) \,_{\pm
}\phi _{p_{3},m,l,\sigma }\left( x,x^{\prime }\right) ,\;l\geq 1\,,  \notag
\\
&&\sum_{\sigma =\pm 1}\,_{\pm }\Psi _{p_{3},m_{+},l,-\sigma }\left( x\right)
\,_{\pm }\overline{\Psi }_{p_{3},m_{+},l,-\sigma }\left( x^{\prime }\right)
\notag \\
&=&\sum_{\sigma =\pm 1}\frac{1}{2\,_{\pm }\varepsilon }\left( \gamma ^{\nu
}P_{\nu }+M\right) \frac{1}{2}\left( 1+\sigma \Sigma ^{3}\right) \,_{\pm
}\phi _{p_{3},m_{+},l,-\sigma }\left( x,x^{\prime }\right) ,\;l\leq 0\,,
\label{g33}
\end{eqnarray}
and for $\omega =0$, we have
\begin{equation*}
\,_{\pm }\Psi _{p_{3},0,l,-1}\left( x\right) \,_{\pm }\overline{\Psi }%
_{p_{3},0,l,-1}\left( x^{\prime }\right) =\frac{1}{2\,_{\pm }\varepsilon }%
\left( \gamma ^{\nu }P_{\nu }+M\right) \frac{1}{2}\left( 1-\Sigma
^{3}\right) \,_{\pm }\phi _{p_{3},0,l,-1}\left( x,x^{\prime }\right) \,,
\end{equation*}
where
\begin{equation}
\,_{\pm }\phi _{p_{3},m,l,\sigma }\left( x,x^{\prime }\right) =\frac{1}{2\pi
}e^{-i\,_{\pm }\varepsilon \Delta x^{0}-ip_{3}\Delta x^{3}}\phi _{m,l,\sigma
}\left( x_{\bot },x_{\bot }^{\prime }\right) ,\;\Delta x^{3}=x^{3}-x^{\prime
3}\,.  \label{g33a}
\end{equation}
The functions $\phi _{m,l,\sigma }\left( x_{\bot },x_{\bot }^{\prime
}\right) $\ are defined in (\ref{g13}). Therefore,
\begin{eqnarray}
&&S^{c}\left( x,x^{\prime }\right) =\left( \gamma ^{\nu }P_{\nu }+M\right)
\Delta ^{c}\left( x,x^{\prime }\right) \,,  \notag \\
&&\Delta ^{c}\left( x,x^{\prime }\right) =i\sum_{m,l,\sigma }\int_{-\infty
}^{\infty }dp_{3}\frac{1}{2}\left( 1+\sigma \Sigma ^{3}\right)  \notag \\
&&\times \left[ \theta \left( \Delta x^{0}\right) \frac{1}{%
2\,_{+}\varepsilon }\,_{+}\phi _{p_{3},m,l,\sigma }\left( x,x^{\prime
}\right) -\theta \left( -\Delta x^{0}\right) \frac{1}{2\,_{-}\varepsilon }%
\,_{-}\phi _{p_{3},m,l,\sigma }\left( x,x^{\prime }\right) \right] \,.
\label{g34}
\end{eqnarray}
Applying the relations (\ref{g15}), (\ref{g16}), one obtains the proper time
integral representation for $\Delta ^{c}$,
\begin{eqnarray}
&&\Delta ^{c}\left( x,x^{\prime }\right) =\int_{0}^{\infty }f\left(
x,x^{\prime },s\right) ds,\;f\left( x,x^{\prime },s\right) =\sum_{l=-\infty
}^{\infty }f_{l}\left( x,x^{\prime },s\right) \,,  \notag \\
&&f_{l}\left( x,x^{\prime },s\right) =D\left( s\right) \sum_{\sigma =\pm
1}\Phi _{l,\sigma }\left( s\right) e^{-i\sigma eBs}\frac{1}{2}\left(
1+\sigma \Sigma ^{3}\right) \,,  \notag \\
&&D\left( s\right) =\frac{eB}{16\pi ^{2}s\sin \left( eBs\right) }\exp
\left\{ \frac{i}{4s}\left[ \left( \Delta x_{3}\right) ^{2}-\left( \Delta
x_{0}\right) ^{2}\right] -iM^{2}s\right\}  \notag \\
&&\times \exp \left\{ -il_{0}\Delta \varphi +\frac{ieB}{4}\left(
r^{2}+r^{\prime 2}\right) \cot \left( eBs\right) \right\} \,,  \label{g35}
\end{eqnarray}
where $\Phi _{l,\sigma }\left( s\right) $ are defined in (\ref{g25a}).

Carrying out similar calculations for $B<0$ one can verify that (\ref{g35})
is valid for both signs of $B$. Therefore, for any sign of $B,$ we get
\begin{eqnarray}
&&f_{nc}\left( x,x^{\prime },s\right) =\sum_{l\neq 0}f_{l}\left( x,x^{\prime
},s\right)  \notag \\
&=&D\left( s\right) e^{-ieBs\left( \mu +\Sigma ^{3}\right) }\left\{ Y\left(
z,\Delta \varphi -eBs,\mu \right) +Y\left( z,-\Delta \varphi +eBs,-\mu
\right) \right.  \notag \\
&&\left. +\left[ e^{-\frac{i\pi \mu }{2}}J_{\mu }\left( z\right)
-e^{-i\left( \Delta \varphi -eBs\right) }e^{-\frac{i\pi \left( 1-\mu \right)
}{2}}J_{1-\mu }\left( z\right) \right] \frac{1}{2}\left( 1+\Sigma
^{3}\right) \right\} \,,  \notag \\
&&f_{0}^{\left( -\pi /2\right) }\left( x,x^{\prime },s\right) =\frac{1}{2}%
D\left( s\right) e^{-i\mu eBs}\left[ e^{-i\Delta \varphi }e^{-\frac{i\pi
\left( 1-\mu \right) }{2}}J_{1-\mu }\left( z\right) \left( 1+\Sigma
^{3}\right) \right.  \notag \\
&&\left. +e^{ieBs}e^{\frac{i\pi \mu }{2}}J_{-\mu }\left( z\right) \left(
1-\Sigma ^{3}\right) \right] \,,  \notag \\
&&f^{\left( -\pi /2\right) }\left( x,x^{\prime },s\right) =f_{nc}\left(
x,x^{\prime },s\right) +f_{0}^{\left( -\pi /2\right) }\left( x,x^{\prime
},s\right) \,.  \label{g38}
\end{eqnarray}
Using the corresponding solutions for the case $\Theta =\pi /2,$ we obtain
\begin{eqnarray}
&&f_{0}^{\left( \pi /2\right) }\left( x,x^{\prime },s\right) =\frac{1}{2}%
D\left( s\right) e^{-i\mu eBs}\left[ e^{-i\Delta \varphi }e^{-\frac{i\pi
\left( \mu -1\right) }{2}}J_{\mu -1}\left( z\right) \left( 1+\Sigma
^{3}\right) \right.  \notag \\
&&\left. +e^{ieBs}e^{-\frac{i\pi \mu }{2}}J_{\mu }\left( z\right) \left(
1-\Sigma ^{3}\right) \right] \,,  \notag \\
&&f^{\left( \pi /2\right) }\left( x,x^{\prime },s\right) =f_{nc}\left(
x,x^{\prime },s\right) +f_{0}^{\left( \pi /2\right) }\left( x,x^{\prime
},s\right) \,.  \label{g39}
\end{eqnarray}

\section{Summary}

Various Green functions of the Dirac equation with the magnetic-solenoid
field are constructed as sums over exact solutions of this equation. We
stress that doing that we had to take into account all the peculiarities
related to the self-adjoint extension problem of the Dirac operator in the
background under consideration. Both $2+1$ and $3+1$ dimensional cases are
considered. Compact form for the Green functions was obtained thanks to the
important relations (\ref{g12}) and (\ref{g33}) derived by us for the exact
solutions under consideration. The representations of the Green functions as
proper time integrals are constructed. The kernels of the proper time
integrals are represented both as infinite sums over the orbital quantum
number $l$ and as simple integrals. The Green functions are obtained for two
natural self-adjoint extensions, one for the positive solenoid flux and the
other one for the negative solenoid flux. The physical motivation for the
choice of these extensions is their correspondence to the presence of the
point-like magnetic field at the origin and their close relation to the MIT
boundary conditions \cite{BFKS00,BFS99,FK01}. Thus, the considered cases are
of most interest for applications. Other values of the extension parameter
correspond to additional contact interactions \cite{MT93}, and some of the
values are of physical interest as well. To find a closed form of Green
functions for the arbitrary value of the extension parameter is a more
complicated task. The spectra of the corresponding extensions in the
critical subspace are no longer periodic for such a situation that requires
to apply more exquisite calculation methods. We suppose to consider this
issue in our future publications.

In addition, the nonrelativistic Green functions are constructed. The latter
Green functions are represented for all possible types of $2+1$ dimensional
nonrelativistic particles.

\section{Acknowledgments}

D.M.G. is grateful to the Erwin Schr\"{o}dinger Institute for hospitality
and to the foundations CNPq and FAPESP for permanent support. A.A.S. and
S.P.G thank FAPESP for support.

\section{Appendix}

For the sake of completeness we consider here the Green functions for the
scalar particle. They are defined by Eqs. (\ref{g1}), (\ref{g2}), (\ref{g7}%
), and (\ref{g8a}), where $S^{\mp }\left( x,x^{\prime }\right) $ read
\begin{equation*}
S^{\mp }\left( x,x^{\prime }\right) =\pm i\sum \,_{\pm }\phi _{n}\left(
x\right) \,_{\pm }\phi _{n}^{\ast }\left( x^{\prime }\right) \,,
\end{equation*}
and $_{\pm }\phi _{n}\left( x\right) $ form a complete set of
orthonormalized solutions of the Klein-Gordon equation. Here we consider the
natural extension of the Klein-Gordon operator for which solutions in $2+1$
dimensions and the related spectrum read \cite{GGS03a},
\begin{eqnarray*}
&&\,_{\pm }\phi _{m,l}\left( x\right) =\frac{1}{\sqrt{2\varepsilon }}%
e^{-i\,_{\pm }\varepsilon x^{0}}\sqrt{\frac{\gamma }{2\pi }}e^{i\left(
l-l_{0}\right) \varphi }I_{m+\left| l+\mu \right| ,m}\left( \rho \right) \,,
\\
&&\,_{\pm }\varepsilon =\pm \sqrt{M^{2}+\omega },\;\omega =\gamma \left[
1+2m+\left| l+\mu \right| +\xi \left( l+\mu \right) \right] \,, \\
&&l=0,\pm 1,\pm 2,...,\;m=0,1,2,...\,.
\end{eqnarray*}
Using Eqs. (\ref{g15}), (\ref{g16}), (\ref{g16a}), and (\ref{g24}), we
calculate the causal and anticausal propagators. They have the form
\begin{eqnarray}
&&S^{c}\left( x,x^{\prime }\right) =\int_{0}^{\infty }f^{sc}\left(
x,x^{\prime },s\right) ds,\;S^{\bar{c}}\left( x,x^{\prime }\right)
=\int_{-0}^{-\infty }f^{sc}\left( x,x^{\prime },s\right) ds\,,  \notag \\
&&f^{sc}\left( x,x^{\prime },s\right) =\sum_{l}f_{l}^{sc}\left( x,x^{\prime
},s\right) \,,\;f_{l}^{sc}\left( x,x^{\prime },s\right) =A\left( s\right)
e^{il\Delta \varphi }e^{-i\left( l+\mu \right) eBs}e^{-\frac{i\pi \left|
l+\mu \right| }{2}}J_{\left| l+\mu \right| }\,,  \notag \\
&&f^{sc}\left( x,x^{\prime },s\right) =A\left( s\right) e^{-i\mu eBs}\left[
e^{-\frac{i\pi \mu }{2}}J_{\mu }\left( z\right) \right.  \notag \\
&&\left. +Y\left( z,\Delta \varphi -eBs,\mu \right) +Y\left( z,-\Delta
\varphi +eBs,-\mu \right) \right] \,,  \label{a2}
\end{eqnarray}
where $A\left( s\right) $\ is given in (\ref{g250}), and $Y\left( z,\eta
,\mu \right) $ in (\ref{y-intro}), (\ref{fsum}). The expression (\ref{a2})
can be generalized for the $\left( D+1\right) $-dimensional case, where $D$
is the number of spacial dimensions, with the substitution $A\left( s\right)
$ in (\ref{a2}) by $A^{\left( D\right) }\left( s\right) $,

\begin{equation*}
A^{\left( D\right) }\left( s\right) =A\left( s\right) \exp \left\{ \frac{i}{%
4s}\sum_{k=3}^{D}\left( \Delta x_{k}\right) ^{2}\right\} \left( \frac{e^{-%
\frac{i\pi }{2}}}{4\pi s}\right) ^{\left( D-2\right) /2},\;D\geq 3\,.
\end{equation*}

\end{document}